\documentclass[sigconf,screen,letterpaper]{acmart}

\usepackage{booktabs}   %% For formal tables:
\usepackage{subcaption} %% For complex figures with subfigures/subcaptions
\usepackage{color}
\usepackage[textwidth=3.1cm]{todonotes}
\usepackage{xspace}
\usepackage[frozencache]{minted}
\usepackage{cleveref}
\usepackage{mathpartir}
\usepackage{etoolbox}
\usepackage{tikz}
\usepackage{amsmath}
\usepackage{wrapfig}
\usepackage{multirow}
\usepackage{microtype}
\usepackage{graphicx}
\graphicspath{ {./results} }

\definecolor{mike}{rgb}{0.91, 0.84, 0.42}
\definecolor{cdiss}{HTML}{6ED7F7}
\definecolor{awells}{HTML}{ff3377}
\definecolor{shaobo}{HTML}{33FFBD}
\definecolor{aeline}{HTML}{22c442}
\definecolor{khieta}{HTML}{fccbec}

\newtoggle{todos}
\togglefalse{todos}

\iftoggle{todos}{
  \newcommand{\Todo}[3]{\todo[color=#1,size=\tiny]{#2: #3}}
}{
  \newcommand{\Todo}[3]{}
}
\newcommand{\mike}[1]{\Todo{mike}{Mike}{#1}}

\newcommand{\shaobo}[1]{\Todo{shaobo}{Shaobo}{#1}}

\newcommand{\khieta}[1]{\Todo{khieta}{Kesha}{#1}}

\iftoggle{todos}{
  \newcommand{\TODO}[1]{{\color{red}TODO: #1}}
}{
  \newcommand{\TODO}[1]{}
}
\iftoggle{todos}{
  
}{
  
}
\iftoggle{todos}{
  \newcommand{\tocite}[1]{{\color{red}[tocite: #1]}}
}{
  \newcommand{\tocite}[1]{}
}
\newcommand{\code}[1]{%
  \mintinline[fontsize=\small{},mathescape,escapeinside=@@]{go}{#1}%
}

\newminted[cedarblock]{go}{fontsize=\small{},mathescape,breaklines,escapeinside=@@}
\newminted[schemablock]{lisp}{fontsize=\footnotesize{},mathescape,breaklines,escapeinside=@@}

\newcommand{\kw}[1]{\ensuremath{\mathtt{#1}}}

\newtoggle{blind}
\togglefalse{blind}

\iftoggle{blind}{
  \newcommand{\cedar}{Steel\xspace}
}{
  \newcommand{\cedar}{Cedar\xspace}
}

\newcommand{\tighten}{\looseness=-1}

\usepackage[english]{babel}
\addto\extrasenglish{
  
}

\setcopyright{rightsretained}
\acmDOI{10.1145/3663529.3663854}
\acmYear{2024}
\copyrightyear{2024}
\acmSubmissionID{fsecomp24industry-p88-p}
\acmISBN{979-8-4007-0658-5/24/07}
\acmConference[FSE Companion '24]{Companion Proceedings of the 32nd ACM International Conference on the Foundations of Software Engineering}{July 15--19, 2024}{Porto de Galinhas, Brazil}
\acmBooktitle{Companion Proceedings of the 32nd ACM International Conference on the Foundations of Software Engineering (FSE Companion '24), July 15--19, 2024, Porto de Galinhas, Brazil}
\received{2024-02-08}
\received[accepted]{2024-04-18}

\begin{document}

\title{How We Built Cedar: A Verification-Guided Approach}

\author{Craig Disselkoen}
\orcid{0000-0003-4358-2963}
\affiliation{%
  \institution{Amazon Web Services}
  %\city{Arlington}
  \country{USA}
}
%\email{cdiss@amazon.com}

\author{Aaron Eline}
\orcid{0000-0002-9105-4922}
\affiliation{%
  \institution{Amazon Web Services}
  %\city{Arlington}
  \country{USA}
}
%\email{aeline@amazon.com}

\author{Shaobo He}
\orcid{0000-0002-9899-6226}
\affiliation{%
  \institution{Amazon Web Services}
  %\city{Santa Clara}
  \country{USA}
}
%\email{shaobohe@amazon.com}

\author{Kyle Headley}
\orcid{0000-0002-4880-4150}
\affiliation{%
  \institution{Unaffiliated}
  %\city{Arlington}
  \country{USA}
}
%\email{kylenheadley@gmail.com}

\author{Michael Hicks}
\orcid{0000-0002-2759-9223}
\affiliation{%
  \institution{Amazon Web Services}
  %\city{Arlington}
  \country{USA}
}
\affiliation{%
  \institution{University of Maryland}
  %\city{College Park}
  \country{USA}
}
%\email{mwhicks@amazon.com}

\author{Kesha Hietala}
\orcid{0000-0002-2724-0974}
\affiliation{%
  \institution{Amazon Web Services}
  %\city{Arlington}
  \country{USA}
}
%\email{khieta@amazon.com}

\author{John Kastner}
\orcid{0000-0002-1273-5990}
\affiliation{%
  \institution{Amazon Web Services}
  %\city{Arlington}
  \country{USA}
}
%\email{jkastner@amazon.com}

\author{Anwar Mamat}
\orcid{0009-0007-1184-7206}
\affiliation{%
  \institution{University of Maryland}
  %\city{College Park}
  \country{USA}
}
%\email{anwarmamat@gmail.com}

\author{Matt McCutchen}
\orcid{0000-0003-4814-5148}
\affiliation{%
  \institution{Unaffiliated}
  %\city{Arlington}
  \country{USA}
}
%\email{matt@mattmccutchen.net}

\author{Neha Rungta}
\orcid{0000-0001-5143-8940}
\affiliation{%
  \institution{Amazon Web Services}
  %\city{Santa Clara}
  \country{USA}
}
%\email{rungta@amazon.com}

\author{Bhakti Shah}
\orcid{0009-0000-2698-0260}
\affiliation{%
  \institution{University of Chicago}
  %\city{Chicago}
  \country{USA}
}
%\email{bhaktishah@uchicago.edu}

\author{Emina Torlak}
\orcid{0000-0002-1155-2711}
\affiliation{%
  \institution{Amazon Web Services}
  %\city{Seattle}
  \country{USA}
}
%\email{torlaket@amazon.com}

\author{Andrew Wells}
\orcid{0000-0001-7780-2122}
\affiliation{%
  \institution{Amazon Web Services}
  %\city{Santa Clara}
  \country{USA}
}
%\email{anmwells@amazon.com}

%% Abstract
%% Note: \begin{abstract}...\end{abstract} environment must come
%% before \maketitle command

\begin{abstract}
This paper presents \emph{verification-guided development} (VGD), a software engineering process we used to build \cedar, a new policy language for expressive, fast, safe, and analyzable authorization.
Developing a system with VGD involves writing an executable model of the system and mechanically proving properties about the model; writing production code for the system and using \emph{differential random testing} (DRT) to check that the production code matches the model; and using \emph{property-based testing} (PBT) to check properties of unmodeled parts of the production code.
Using VGD for \cedar, we can build fast, idiomatic production code, prove our model correct, and find and fix subtle implementation bugs that evade code reviews and unit testing. While carrying out proofs, we found and fixed 4 bugs in \cedar's policy validator, and DRT and PBT helped us find and fix 21 additional bugs in various parts of \cedar.
\end{abstract}

\begin{CCSXML}
  <ccs2012>
     <concept>
         <concept_id>10002978.10002986</concept_id>
         <concept_desc>Security and privacy~Formal methods and theory of security</concept_desc>
         <concept_significance>500</concept_significance>
         </concept>
     <concept>
         <concept_id>10011007.10011074.10011092</concept_id>
         <concept_desc>Software and its engineering~Software development techniques</concept_desc>
         <concept_significance>500</concept_significance>
         </concept>
     <concept>
         <concept_id>10002978.10002991.10010839</concept_id>
         <concept_desc>Security and privacy~Authorization</concept_desc>
         <concept_significance>500</concept_significance>
         </concept>
   </ccs2012>
\end{CCSXML}
  
  \ccsdesc[500]{Security and privacy~Formal methods and theory of security}
  \ccsdesc[500]{Software and its engineering~Software development techniques}
  \ccsdesc[500]{Security and privacy~Authorization}

\keywords{formal methods, fuzz testing, differential testing, policy authorization language, interactive theorem proving}

\maketitle
\renewcommand{\shortauthors}{Disselkoen et al.}

%% Outline: https://quip-amazon.com/tS0dA5zbNvAn/FSE-Short-Paper-on-Verification-Guided-Development

\section{Introduction}
\label{sec:intro}

Cedar~\cite{cedar,cedar-oopsla} is a new, open-source authorization policy language.
Developers express permissions for their applications as policies written in Cedar.
When the application needs to perform a sensitive user operation, it sends a request to the Cedar authorization engine, which allows or denies the request by evaluating the relevant policies.
Because Cedar policies are separate from application code, they can be independently authored, updated, analyzed, and audited.\tighten

Cedar's authorization engine is part of an application's \emph{trusted computing base} (TCB), which comprises components that are critical to the application's security.
To provide assurance that the engine's authorization decisions are correct, we develop Cedar using a two-part process we call \emph{verification-guided development} (VGD).
First, we construct simple and readable formal models of Cedar's components.
We write these models in the Lean programming language~\cite{moura2021lean}, and carry out mechanized proofs to show that they satisfy important correctness properties.
Second, we use \emph{differential random testing} (DRT)~\cite{mckeeman1998differential} to show that the models match the production code, which is written in Rust.
DRT involves generating millions of inputs---consisting of policies, data, and requests---and testing that the model and production code agree on the outputs.
We also perform \emph{property-based testing} (PBT) of the Rust code directly (in the style of QuickCheck~\cite{claessen2000quickcheck}) when there is no corresponding Lean model.
We may also property-test conjectured properties on the Rust code before we prove them in Lean.\tighten

VGD provides a practical balance of competing concerns around high assurance, ease of development, and maintainability.
To see why, consider two other processes we might have followed.

One approach would be to develop Cedar entirely in Lean, compile to C, and deploy the generated C code in production.
In addition to the formal model, we would write an optimized and full-featured implementation (which handles errors and messages more carefully, provides parsers for various formats, etc.) in Lean, and then prove the two equivalent.
A key benefit of this approach is that we would \emph{formally verify}---rather than just test---the equivalence of the deployed code and model.
But there are significant downsides to writing production applications in Lean and deploying its generated C code.
Lean is a new programming language, so it has a limited developer pool and lacks useful libraries available in mainstream languages.
Debugging a failure might necessitate stepping through the generated C code and mapping it back to the Lean source, requiring expertise in C and Lean.
Doing so could be particularly difficult if the failure is due to an interaction between handwritten code and Lean-generated C code.
In addition, C is a memory-unsafe language, so a bug in the Lean compiler could lead to security issues.\tighten

Another approach would be to develop only in Rust and prove correctness of the Rust code directly, using a tool like Aeneas~\cite{ho2022aeneas}, Kani~\cite{kani}, Prusti~\cite{astrauskas2022prusti}, Creusot~\cite{denis2022creusot}, or Verus~\cite{lattuada2023verus}.
The main challenge here is that tools for verifying Rust code are not (yet) up to the task, e.g., they cannot handle much of the standard library, they support only certain code idioms, they sometimes have trouble scaling, and they are limited in the properties one can specify.
The model consumable by one of these tools is unlikely to be as readable as the Lean model, due to Rust's low-level nature.
These limitations present challenges to guaranteeing high assurance, but also to ease of development and maintainability, because teams would likely need to wrangle code into less idiomatic forms with each release just to enable proofs.\tighten

Using VGD has proved beneficial for \cedar.
It has helped us improve \cedar's design: While implementing the formal model and carrying out proofs of soundness of the \cedar policy validator, we found and fixed four bugs.
It has also allowed us to write fast, idiomatic Rust code with increased confidence in its correctness: Using DRT and PBT we have so far found and fixed 21 bugs in various \cedar components.

We believe VGD represents a practical approach to leveraging the benefits of formal methods while also assuring the deployed code is easy to use, develop, and maintain.
The remainder of this paper presents our experience with VGD for \cedar in detail, beginning with background on \cedar (\Cref{sec:cedar}), our Lean models of and proofs about \cedar's components (\Cref{sec:models}), how we use DRT and PBT on our Rust code (\Cref{sec:drt}), and how VGD compares to related work (\Cref{sec:related}).

Cedar is open source.
Our Lean models, Rust implementation, and testing setup are all available at \url{https://github.com/cedar-policy}. 
%\section{Background}
%
\section{The Cedar Policy Language}
\label{sec:cedar}

Cedar is a language for writing authorization policies, supporting idioms in the style of role-based access control (RBAC)~\cite{rbac}, attribute-based access control (ABAC)~\cite{abac}, and their combination.
Cedar policies use a syntax resembling natural language to define who (the principal) can do what (the action) on which target (the resource) under what conditions.
To see how Cedar works, consider a simple application, TinyTodo~\cite{tinytodo}, designed for managing task lists.
TinyTodo uses Cedar to control who can do what.

\begin{figure}
    \centering
    \begin{minipage}{.75\columnwidth}
    \begin{cedarblock}
    // Policy 1
    permit(principal, action, resource)
    when {
        resource has owner && 
        resource.owner == principal
    };

    // Policy 2
    permit(
        principal,
        action == Action::"GetList",
        resource)
    when {
        principal in resource.readers || 
        principal in resource.editors
    };

    // Policy 3
    forbid (
        principal in Team::"interns",
        action == Action::"CreateList",
        resource == Application::"TinyTodo"
    );
    \end{cedarblock}
    \end{minipage}
    \caption{\cedar policies for TinyTodo}
    \label{fig:tinytodo-policies}
    \end{figure}

\Cref{fig:tinytodo-policies} shows three sample policies from TinyTodo.
The first is an ABAC-style policy that allows any principal (a TinyTodo user) to perform any action on a resource (a TinyTodo list) they own, as defined by the resource's \code{owner} attribute matching the requesting principal.
The second policy allows a principal to read the contents of a task list (\code{Action::"GetList"}) if the principal is in the list's \code{readers} or \code{editors} group.
The third is an RBAC-style policy that forbids interns (a role) from creating a new task list (\code{Action::"CreateList"}) using TinyTodo (\code{Application::"TinyTodo"}).\tighten

When the application needs to enforce access, as when a user of TinyTodo issues a command, it makes a corresponding request to the \textbf{Cedar authorizer}.
The authorizer invokes the \textbf{Cedar evaluator} for each policy, to see whether it is satisfied by the request (and provided application data).
The authorizer returns decision \emph{Deny} if no \code{permit} policy is satisfied or if any \code{forbid} policy is satisfied.
It returns \emph{Allow} if at least one \code{permit} policy is satisfied and no \code{forbid} policies are.
Cedar supports indexing policies so that they be can quickly \emph{sliced} to a subset relevant to a particular request.
For example, if a given request does not have \code{Application::"TinyTodo"} as its resource, then Policy 3 is not included in the slice.\tighten

Principals, resources, and actions are \cedar \emph{entities}.
Entities are collected in an \emph{entity store} and referenced by a unique identifier consisting of the \emph{entity type} and \emph{entity ID}, separated by "\code{::}".
Each entity is associated with zero or more attributes mapped to values, and zero or more parent entities.
The parent relation on entities forms a directed acyclic graph (DAG), called the \emph{entity hierarchy}.
Expression \code{A in B} evaluates to true if \code{B} is a (transitive) parent of \code{A}.\tighten

\cedar policies have three components: the \emph{effect}, the \emph{scope}, and the \emph{conditions}.
The effect is either \code{permit} or \code{forbid}, indicating whether the policy is granting or removing access.
The scope comes after the effect, constraining the principal, action, and resource components of a request; policy indexing is based on scope constraints.
The conditions are optional expressions that come last, adding further constraints, oftentimes based on attributes of request components.
Policies access request components using the variables \code{principal}, \code{action}, \code{resource}, and \code{context}.\tighten

%ReBAC-style policies use attributes that reference entity groups to express \emph{relations} between two entities.
%For example, Policy 3 is a ReBAC-style policy, which states that any principal can read the contents of a task list (\code{Action::"GetList"}) if they are in either the list's \code{readers} or \code{editors} groups.
%Here, \code{principal in resource.readers} and \code{principal in resource.editors} are expressions that can be viewed as querying whether \code{principal} is in the \emph{readers} and \emph{editors} relations with \code{resource}.
%
%For our running example, suppose \code{User::"andrew"} shares \code{List::"0"} with the team \code{interns} as a reader (which he is allowed to do because of Policy 1).
%As a result, TinyTodo will update the hierarchy in \Cref{fig:hierarchy-example} to add a parent edge from \code{Team::"interns"} to \code{Team::"1"}.
%Now if \code{User::"aaron"} attempts to read the contents of \code{List::"0"} he will be allowed to do so according to Policy 3: its condition \code{principal in resource.readers} is true since the \code{readers} attribute of \code{List::"0"} corresponds to \code{Team::"1"}, which is an ancestor of \code{User::"aaron"} in the entity hierarchy.

Policy evaluation could result in a dynamic type error, e.g., if a \code{when} expression tries to access \code{resource.pwner} but \code{resource} has no \code{pwner} attribute, or tries to use a numeric operation like \code{<} on a pair of entities.
When this happens, the erroring policy does not factor into the final authorization decision---it is ignored.
Users can avoid this situation by using the \textbf{Cedar validator} to statically check their policies against a \emph{schema}, which defines the names, shapes, and parent relations of entity types, as well as the legal actions upon them.
If the validator is satisfied, users can be sure that when requests conform to the schema, their policies' evaluation will never result in run-time type errors.\tighten

% \Cref{fig:tinytodo-schema} shows a fragment of TinyTodo schema.

% \begin{figure}
%     \centering
%     \begin{schemablock}
%     entity Application;
%     entity Team, User in [Team, Application];
%     entity List in [Application] {
%         readers: Team,
%         editors: Team,
%         owner: User
%     };
%     action CreateList appliesTo {
%         principal: [User],
%         resource: [Application]
%     };
%     action GetList appliesTo { principal: [User], resource: [List] };
%     \end{schemablock}
%     \caption{TinyTodo schema}
%     \label{fig:tinytodo-schema}
%     \end{figure}

%\todo{Slicing}
%\todo{Validation}

To help users understand the meaning of their policies, we designed Cedar to be amenable to \emph{automated reasoning} via an encoding of its semantics into formal logic.
The \textbf{Cedar symbolic compiler} translates Cedar policies to SMT-lib~\cite{smtlib}, the language of SMT solvers, producing an encoding that is sound, complete, and decidable.
With this compiler we can, for example, prove that two sets of policies are equivalent, meaning that they authorize exactly the same requests in the same entity stores.
To do this, we encode each policy set as a formula and ask the SMT solver to search for a request and entity store that is allowed by one policy set but not the other.
A response of \emph{UNSAT} guarantees that the policy sets are equivalent.\tighten

%\subsection{Rigorous Reasoning about Cedar's Correctness}
%\todo{Talk about Lean}

%\emph{[Baseline content, taken from Emina's blog post]}
%We build a formal model of Cedar in Lean~\cite{moura2021lean}, and use Lean's proof assistant to prove that Cedar's components satisfy key safety and security properties.
%Lean is a full-featured functional programming language, so writing Cedar models amounts to writing high-level, executable prototypes of each core component.
%Lean is also very fast, allowing us to efficiently test our production code (written in Rust) against the Lean models.
%Until now, Lean has been used mostly by mathematicians to formalize complex math proofs~\cite{hartnett_effort_2020}.
%But our experience shows it is also an excellent tool for proofs about software correctness and security.
%
%We adopted Lean to write our model because of its fast runtime, extensive libraries, and small trusted computing base (TCB).
%The fast runtime enables efficient differential testing of Cedar models.
%The libraries provide reusable verified data structures and tactics built by the open-source community.
%And Lean's small TCB lets us leverage these contributions without having to trust that they are correct---Lean checks that they are, and we only need to trust Lean's small proof checking kernel.

%\subsection{Random Testing}
%\todo{Talk about PBT and fuzzing}
\section{Lean Models and Proofs}
\label{sec:models}

The first part of the verification-guided development process is constructing an executable, formal model of the system, using a proof-oriented language.
We wrote models of the Cedar evaluator, authorizer, and validator in Lean~\cite{moura2021lean}, and are in the process of writing one for the symbolic compiler.
The models serve as Cedar's \emph{specification}, and we had two goals when writing them.
First, the models should be \emph{human readable}, and thus favor simplicity and understandability (\Cref{sec:spec}).
Second, they should be as \emph{feature complete} as possible, so that proofs about them (\Cref{sec:properties}) apply to the full language, not just an abstraction, and so that the models can be used as oracles for differential testing (\Cref{sec:drt}).

\subsection{A human-readable specification}
\label{sec:spec}

Lean allows us to write concise specifications.
As an illustration, here's the Lean code for the Cedar authorizer:
\begin{cedarblock}
def isAuthorized (req : Request) (entities : Entities) 
                 (policies : Policies) : Response :=
  let forbids := 
      satisfiedPolicies .forbid policies req entities
  let permits := 
      satisfiedPolicies .permit policies req entities
  if forbids.isEmpty && !permits.isEmpty
  then { decision := .allow, policies := permits }
  else { decision := .deny,  policies := forbids }
\end{cedarblock}
This code naturally expresses the logic that an authorization decision is allowed as long as some permit policy is satisfied and no forbid policies are.

Lean is a purely functional language with algebraic data types, so it was easy to directly express Cedar's evaluator, validator, and symbolic compiler as recursive functions over abstract syntax trees.
That said, Lean requires all recursive definitions to be well-founded (so functions always terminate), which complicates modeling of complex recursive structures.
While Rust lets us represent Cedar values as a recursive datatype over built-in sets and maps, Lean prohibits doing so for its standard set and map datatypes because their invariants introduce circular (not well-founded) reasoning.
We worked around this by developing custom set and map datatypes that replace embedded invariants with separate theorems.
Pleasantly, our termination proofs turned out to not be merely tedium: Constructing them helped us uncover and fix a non-termination bug in our definition of the Cedar validator, which would have been difficult to detect through testing.
We also found three other bugs while attempting to carry out the validator soundness proof.
\TODO{We should highlight these in the appendix. They are in the original trophy case Quip.}

\Cref{tab:lean-and-rust} compares the size of the Lean specification and proofs against the corresponding Rust implementation and tests, measured in lines of code (LOC).
The Lean models are an order of magnitude smaller than their Rust counterparts, which include optimizations, extra code to provide useful diagnostic output, and some unmodeled code.
For example, our parsers are not modeled in Lean because there is currently no library support for parser generators, and it is unclear what properties we could prove about a parser model.
We also did not port our unit tests to Lean because we can instead \emph{prove} properties of interest.

\begin{table}[tb]
    \centering
    \caption{Lean and Rust implementations; numbers in LOC}
    \label{tab:lean-and-rust}
    \begin{tabular}{|p{0.21\columnwidth}|p{0.1\columnwidth}|p{0.1\columnwidth}|p{0.1\columnwidth}|p{0.1\columnwidth}|p{0.1\columnwidth}|}
    \hline
        \textbf{Component}& \textbf{Lean model} & \textbf{Lean proofs}  & \textbf{Rust prod}  & \textbf{Rust tests} & \textbf{Rust other} \\\hline
        Custom sets and maps & 244 & 681 & n/a & n/a  & n/a  \\\hline
        Parser & n/a & n/a & 4114 & 3599 & n/a   \\\hline
        Evaluator and Authorizer &	897	& 347 &	4877 & 7061 & n/a \\\hline
        Validator &	532 & 4686 & 6702 & 9798 & n/a \\\hline
        \textbf{Total} & 1673 & 5714 & 15693 & 20458 & 31391 \\\hline
    \end{tabular}
\end{table}

\subsection{Proofs of properties}
\label{sec:properties}

We used Lean to formalize and prove several properties of our Cedar models, listed below.

\begin{enumerate}
\item \label{proof:forbid-trumps-permit} \textbf{Forbid trumps permit}: If any \kw{forbid} policy is satisfied, the request is denied.

\item \label{proof:default-deny} \textbf{Default deny}: If no \kw{permit} policy is satisfied, the request is denied.

\item \label{proof:explicit-allow} \textbf{Explicit allow}: If a request is allowed, some \kw{permit} policy was satisfied.

\item \label{proof:order-independence} \textbf{Order independence}: The authorizer outputs the same decision regardless of policy evaluation order or duplicates.

\item \label{proof:sound-slicing} \textbf{Sound slicing}: The policy slicing algorithm selects a \emph{slice} (i.e., subset) of policies that produces the same authorization decision as the full policy set for a given request and entities.

\item \label{proof:validation-soundness} \textbf{Validation soundness}: If the validator accepts a policy, its evaluation will never result in a type error.

\item \label{proof:termination} \textbf{Termination} Cedar functions always terminate.
\end{enumerate}

Properties~\ref{proof:forbid-trumps-permit}--\ref{proof:order-independence} capture how authorization decisions are made.
Their proofs give us a simple, declarative specification of Cedar's authorization behavior, complementing the executable specification provided by the model.
For example, here is the full Lean statement and proof of Property~\ref{proof:forbid-trumps-permit}:

\newenvironment{alltt}{\ttfamily}{\par}

\begin{small}
\begin{alltt}
\begin{tabbing}
AA\=BB\=CCCC\=DD\kill
theorem forbid\_trumps\_permit (request : Request) \\
\>  (entities : Entities) (policies : Policies) :\\
\>  ($\exists$ (policy : Policy),\\
\>\>    policy $\in$ policies $\wedge$\\
\>\>    policy.effect = forbid $\wedge$\\
\>\>    satisfied policy request entities) $\rightarrow$\\
\>  (isAuthorized request entities policies).decision = deny \\
:= by \\
\>  intro h \\
\>  unfold isAuthorized\\
\>  simp [if\_satisfied\_then\_satisfiedPolicies\_non\_empty \\
\>\>\> forbid policies request entities h]
\end{tabbing}
\end{alltt}
\end{small}
Property~\ref{proof:sound-slicing} shows that the slicing algorithm can be used safely to scale authorization.
Property~\ref{proof:validation-soundness} ensures that Cedar's type system is sound: well-typed Cedar policies cannot ``go wrong.''
This is the most involved proof we have done so far.
Finally, Property~\ref{proof:termination} (in conjunction with the others) guarantees total correctness of Cedar.

Our total proof-to-model ratio is roughly $3.4\!:\!1$ (see \Cref{tab:lean-and-rust}).
Throughout development, we benefited from Lean's extensive library of theorems and its IDE extension, which checks proofs interactively and provides instant feedback.
Our Lean proofs are fast to verify, and the models are fast to execute.
It takes about 3 minutes to check all proofs and compile models for execution.
During differential testing, the median execution time for the Lean authorizer is 6 microseconds, compared to 10 microseconds for Rust.
\khieta{Will leave these numbers as-is for internal review, but I'll have revised numbers supported by more data before the FSE deadline}

% Emina:  I think we should drop its paragraph because it's not an
% issue with Lean. It's an issue with how we wrote the code, and
% we can write a better wildcard-matching
% algorithm in Lean too.  (We did in Dafny):
% The only performance issue of Lean models that we encountered is wildcard matching. We implement it in Lean as a simple recursive function that backtracks and has exponential complexity, whereas the Rust production code uses dynamic programming and thus has polynomial complexity. As a result, we have to limit input sizes of wildcard matching during DRT to avoid timeouts.

\begin{figure}[tb]
    \centering
    \includegraphics[width=\columnwidth]{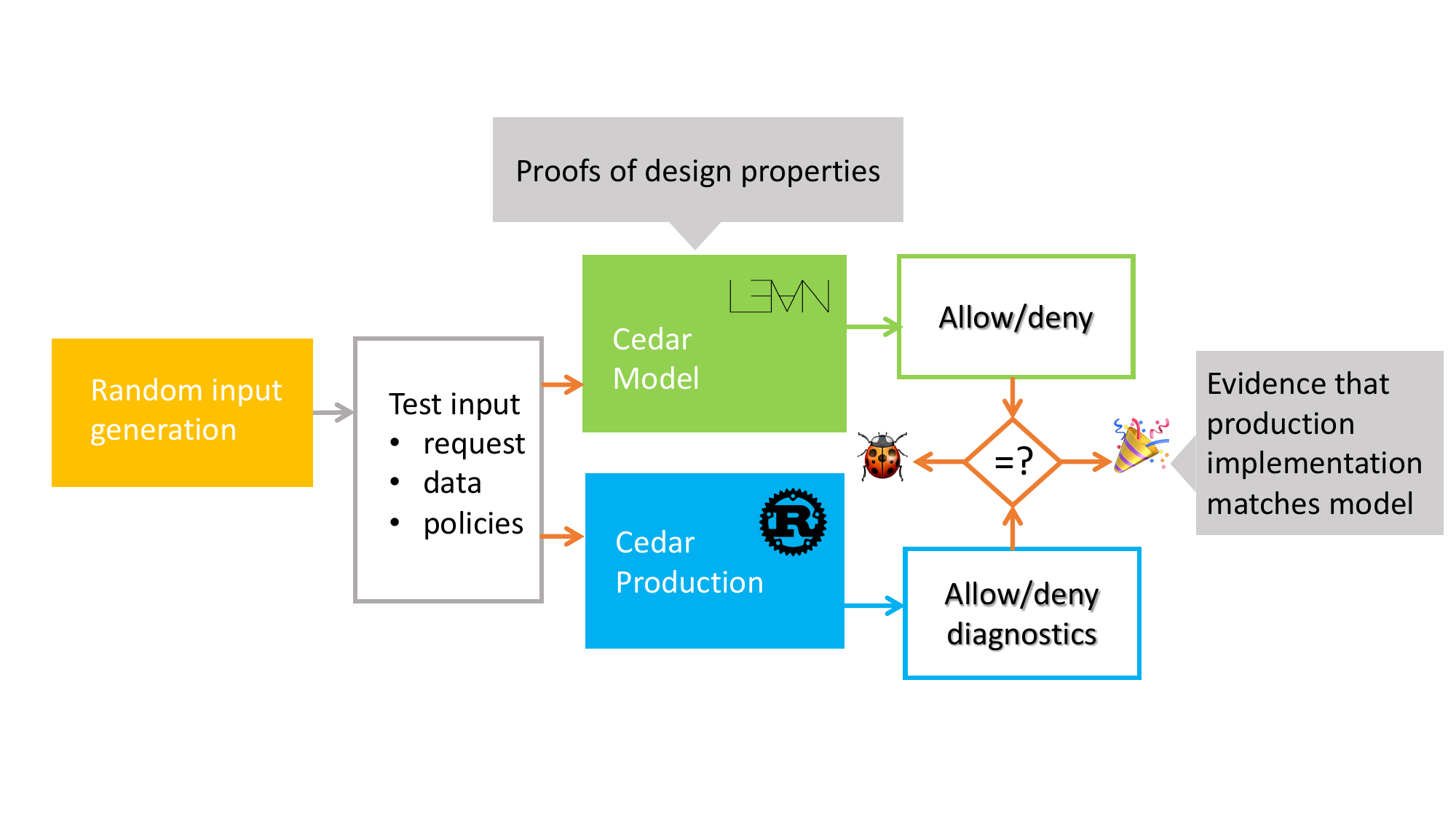}
    \caption{DRT workflow \TODO{Lots of whitespace -- trim to save space}}
    \label{fig:drt-flow}
    \end{figure}

\section{Differential Random Testing}
\label{sec:drt}

\newcommand{\cargofuzz}{\texttt{cargo-fuzz}\xspace}

The second part of the verification-guided development process is to use \emph{differential random testing}~\cite{mckeeman1998differential} to increase our confidence that the behavior of our formal models matches that of our production code.
\Cref{fig:drt-flow} shows the workflow of DRT. %\TODO{Change the picture so that "test input" is not filled in (white background) while "random input generation" is filled in orange. This makes the coloring consistent with the other parts: input/output values are not colored, while components that produce those inputs/outputs are colored.}
Using the \cargofuzz fuzz testing framework~\cite{cargo-fuzz}, we randomly generate millions of inputs---access requests, entities, and policies---and send them to both the Lean model and the corresponding Rust production implementation.
If the two versions agree on the output, then we obtain a piece of evidence that Rust production code is on par with the Lean model.
If they disagree, then we have found a bug (e.g., production code incorrectly implements the specification).
With each version of Cedar, we save a minimized set of \emph{corpus tests} generated by \cargofuzz to use as part of continuous integration testing (\Cref{sec:test-discussion}).

Using the same \cargofuzz framework, we also perform \emph{property-based testing} in the style of QuickCheck~\cite{claessen2000quickcheck} to directly check properties of our production components.
PBT is complementary to DRT because it allows us to test properties we have yet to prove in Lean, and those of production components (such as the Cedar policy parser) for which no model exists.
%For instance, we property-test the soundness of partial evaluation, an experimental feature whose formal model is currently under development.

This section discusses how we generate useful random inputs (\Cref{sec:input-gen}), what properties we test (\Cref{sec:targets}), and what we have learned in the process (\Cref{sec:test-discussion}).

\begin{table*}[tb]
    \caption{Summary of DRT and PBT test targets}
    \centering
    \begin{tabular}{|l|p{0.17\textwidth}|p{0.46\textwidth}|p{0.05\textwidth}|}
    \hline
        \textbf{Property} & \textbf{Input Generator}  & \textbf{Description} & \textbf{\# bugs found} \\ \hline
         & ABAC (type-directed) & Differentially test production authorizer against its Lean model using a \emph{single} \emph{ABAC} policy with a mostly \emph{well-typed} condition & \multirow{2}{*}{6}  \\\cline{2-3}
         \multirow{2}{*}{Rust and Lean authorizer parity} & ABAC & Differentially test production authorizer against its Lean model using a \emph{single} \emph{ABAC} policy with an \emph{arbitrary} condition &   \\ \cline{2-4}
         & RBAC & Differentially test production authorizer against its Lean model using \emph{multiple} \emph{RBAC} (condition-free) policies & 0 \\ \hline
        Rust and Lean validator parity & ABAC (type-directed) & Differentially test production validator against its Lean model & 4 \\ \hline
        Parser roundtrip & ABAC & Test that the composition of pretty-printing and parsing produces a policy identical to the one generated & 6 \\ \hline
        Formatter roundtrip & ABAC & Test that the composition of pretty-printing, formatting, and parsing produces a policy identical to the one generated & 2 \\ \hline
        Parser safety & Random bytes & Simply fuzz the Cedar parser with arbitrary inputs & 0 \\ \hline
        Validation soundness & ABAC (type-directed) & Test that evaluating policies that validate does not produce type errors & 3  \\ \hline
        % SMT encoding concretization & ABAC (Type-Directed)  & Property test the symbolic compiler: SMT encodings with concrete inputs produce the same authorization results as the Cedar evaluator & 6 \\ \hline
    \end{tabular}

    \label{tab:fuzz-targets}
\end{table*}

\subsection{Input generation}
\label{sec:input-gen}

To use DRT effectively requires generating inputs that thoroughly exercise the targeted code.
Na\"ive input generation is insufficient: If we randomly generate a policy, entity store, and request \emph{independently}, policies are likely to be ill-typed and to reference non-existent entities or attributes.
This would over-exercise error-handling code and fail to cover much of a component's core logic.
We tackle this challenge by correlating the generation of policies, entity stores, and requests.

%To resolve this, we wrote several input generators, including ones that take care to generate policies, data, and requests that are consistent with one another, while also producing policies that use Cedar's key language constructs.
Specifically, our typical input generation strategy is \emph{type directed} in the style of \citet{10.1145/1982595.1982615}: we first generate a schema, then an entity store that conforms to the schema, and then policies and requests that access those entities in conformance with the schema.
This approach assures we target the core logic.
But it has the problem that well-typed policies do not exercise error handling code.
Therefore we developed another generator whose policies always refer to entities, actions, and attributes in the schema, but whose \emph{conditions} can be ill-typed.
Because \cargofuzz is a \emph{coverage-guided} testing tool, leveraging \texttt{libfuzzer}~\cite{libfuzzer} to bias input generation toward unexecuted code, we hoped this generator would gravitate to well-typed conditions, too, and we could lean solely on it for input generation.
However, our experience proved otherwise.
The type-directed generator produces more complicated inputs, like set operations, than the non type-directed version, leading to deeper coverage of core logic and faster bug discovery.
So we use both.

\subsection{Properties to test}
\label{sec:targets}

The differential and other properties we test are given in \Cref{tab:fuzz-targets}.
Some properties have multiple testers with different generators.
For example, the property \textit{Rust and Lean authorizer parity} is tested by two different policy generation strategies---ABAC and RBAC. The former produces a single ABAC-style policy for each run and mainly targets the evaluator since generated policies can have nontrivial conditions.
The latter generates multiple RBAC-style policies for each run and aims to exercise the authorization logic that makes a decision involving more than one policy.

%We split authorization decisions into two fragments: firstly the evaluation of policy expressions, and secondly the combining of policy evaluation results to formm an overall authorization decision.
%Thus, we have one test target for each of these two phases.
%The first ("abac") attempts to randomly generate a single policy with a complex \verb|when| clause containing a boolean expression.
%The second ("rbac") generates many simple policies to ensure policy results are combined correctly.

%The main metrics that we use to showcase DRT's efficacy are line coverage and distributions of generated inputs. DRT should cover all lines of Cedar core components such as the evaluator and the validator. However,

\TODO{Include stats about the number of inputs generated, coverage acheived, etc.}

\subsection{Experience}
\label{sec:test-discussion}

We use Amazon's Elastic Container Service (ECS) to test our properties daily on currently supported Cedar versions.
We allocate 4096 CPU units (4 vCPUs) and 8GB memory to fuzz each target for 6 hours.
This setup generates millions of inputs for most targets.
We do not have a particular reason for the 6 hour duration, although it ensures that block coverage for each target saturates or grows slowly near the end of a daily run.
We plan to investigate approaches to choosing an optimal fuzzing duration in the future.

\TODO{It would be nice to talk more about the whole table. It's a little random what we talk about here vs. what's shown in the table.}
As shown in \Cref{tab:fuzz-targets}, differential and property testing have uncovered 21 bugs in total.
As one example, differential testing against the Lean model helped us find a bug in a Rust package we used for parsing IP addresses.
This finding eventually motivated us to write our own IP address parser, replacing the buggy external package.
Some bugs found by differential testing even affected Cedar's language design.
For example, an unreleased early version of Cedar provided a method to get the size of a string.
The model and production code ended up having different implementations (using bytes vs. codepoints vs. graphemes), causing DRT to fail.
We eventually dropped this feature after agreeing that it confuses users more than it benefits them.
Property testing helped us uncover subtle bugs in the Cedar policy parser, in how the formatter handles comments, and how namespace prefixes on application data (e.g., \code{TinyTodo::List::"AliceList"}) are interpreted.\shaobo{what does this bug refer to?} \mike{I think this bug is this one \url{https://sim.amazon.com/issues/CEDAR-180}}
All bugs found by the validation soundness property were found \emph{before} we had completed a soundness proof.

\TODO{To make space to talk about more bugs, I think it would be Ok to cut this paragraph}
Complete line coverage alone does not guarantee effective testing \cite{bohme2022reliability}.
The same lines of code executed with different program state can lead to very different outcomes.
Therefore it is important to test the same or similar code paths with diverse inputs, and to focus on paths of importance.

For Cedar DRT, we found that even when the ABAC policy generator achieves full line coverage, many of the generated inputs exercise error handling code.
Furthermore, 35.5\% of the \texttt{when} condition-expressions generated by the type-directed ABAC policy generator, for a typical DRT run, are boolean literals, which means that one third of policy conditions are trivially true or false and do not trigger interesting code paths.
To address these limitations, we added a new input generator that generates \emph{expressions} of various types (as opposed to just those of Boolean type, as produced by the ABAC policy generators).
%Our experience shows that it produces a better distribution that exercise more interesting code paths.
% \Cref{fig:distro} shows a breakdown inputs produced by the type-directed \emph{policy} generator (first row) and the type-directed \emph{expression} generator (second row).
For the expression generator, only 9.7\% of Boolean-typed expressions are boolean literals.
Oftentimes, such Boolean-type expressions \emph{contain} literals, but not boolean ones---the majority are strings (e.g., as part of \texttt{==} expressions) whose evaluation leads to more code coverage because string handling logic (e.g., unescaping raw strings) is non-trivial and error-prone.

Integration tests made from corpus tests turn out to be a valuable asset for Cedar developers: they helped us quickly catch tricky bugs when developing new features.
For instance, they contain subtle wildcard patterns and Cedar values that exposed bugs in the prototypes of wildcard matching and schema-based parsing.
Integration tests allowed us to fix these bugs before pushing the code, avoiding the delay for the daily DRT run to uncover them.

\subsubsection*{Limitations}

Unsurprisingly, Cedar's testing framework has missed bugs, too.
The most notable example is that it did not discover the non-termination bug described in \Cref{sec:spec}.
The reason for this is that the probability of generating inputs triggering this bug is extremely low.
Our framework is also inherently unable to find bugs outside the scope of the test generators.
For example, we did not discover some parser bugs triggered by malformed policies because our test generators create abstract syntax trees of Cedar policies and thus are limited to produce only syntactically correct policies.
We are investigating grammar-based mutation testing~\cite{10.1145/3605157.3605170} to avoid missing this type of bug in the future.

\Cref{sec:appendix} enumerates all the bugs found, and several missed, by DRT and PBT.

\newcommand{\myparagraph}[1]{\emph{#1}.}

\section{Related Work}
\label{sec:related}

% Authorization languages have been used for over a decade.
% Google Cloud~\cite{gcpiam}, Microsoft Azure~\cite{azurepolicy}, and Amazon Web Services (AWS)~\cite{awsiam} all have proprietary authorization languages.
% These languages differ from Cedar in that they are designed for protecting resources in their respective platforms, while Cedar is designed to be general purpose.
% None of these languages has open source evaluators to which we can compare our development practices.
% Fortunately, there are general purpose languages with open-source implementations.
% Rego is a Datalog-like language with an open source evaluator, \textbf{Open Policy Agent} (OPA)~\cite{opa}.
% It is more expressive than Cedar, but does not offer the same level of analyzability.
% \textbf{Zanzibar}~\cite{zanzibar2019} is a relationship-based access control system developed by Google.
% It is proprietary, but has open source clones such as Auth0 Fine Grained Authorization (FGA)~\cite{openFGA}.
% Cedar supports both attribute-based and relationship-based access control.
% Unlike Cedar, neither the OPA nor the FGA evaluator includes a formal model.
Various communities have developed tools to increase software dependability.
The automotive industry follows the MISRA standards~\cite{misra2023}.
However, these standards aim to avoid common pitfalls rather than guarantee functional correctness.
The FAA requires critical aerospace software have MCDC coverage~\cite{nist2012combinatorial}.
This addresses functional correctness but does not guarantee it.
The highest assurance can be gained by \emph{formally verifying} that deployed software meets a specification and that the specification has particular safety/security properties.
Alternatively, rather than prove software against a specification, we could develop an executable model for the specification and \emph{differentially test} the code against the model, and likewise test that the model has certain properties.
Verification-guided development offers a pleasant compromise: We prove properties about a readable formal model, and rigorously test that the deployed code matches that model.

\myparagraph{Formal methods}
A variety of software, including the CompCert optimizing C compiler~\cite{leroy2016compcert}, the SeL4 microkernel~\cite{10.1145/3378426}, and the EverCrypt cryptography library~\cite{protzenko2020evercrypt}, have been formally verified and deployed in practice.
These systems employ the formal verification frameworks Coq~\cite{COQ2024}, Isabelle/HOL~\cite{NIPKOW2021}, and F$^\star$~\cite{FSTAR2024}, respectively, which center around a domain-specific \emph{proof-oriented programming language} (PPL).
The code from the PPL is either mapped to/from code written in a mainstream language, like C or OCaml, or directly \emph{extracted} (compiled) to it.

Lean is similar to Coq, and in principle we could have deployed our formally verified Lean model of Cedar as extracted C code, rather than building a separate Rust version.
However, as discussed in~\Cref{sec:intro}, such extracted code would be challenging to maintain and operate, e.g., when debugging broader system failures in deployment, because it is not intended to be readable.
To address this issue, EverCrypt deploys readable C code using by a purpose-built idiomatic compiler, KaRaMeL~\cite{protzenko2017verified}, which works on the Low$^\star$ subset of F$^\star$.

Even with such a compiler, developing industrial-strength code in Lean (or indeed, any PPL) is challenging because of its limited library support and limited base of developer expertise, compared to a mainstream language.
Alternatively, one could try to formally verify a software system written in a language like Java (e.g., using tools like OpenJML~\cite{OPENJML2022} or Krakatoa~\cite{FILLIATRE2007}), or Rust (e.g., using tools such as Aeneas~\cite{ho2022aeneas}, Kani~\cite{kani}, Prusti~\cite{astrauskas2022prusti}, Creusot~\cite{denis2022creusot}, or Verus~\cite{lattuada2023verus}).
However, these tools have limitations in scope, scalability, and tooling that prevent their use on an industrial scale.
As Lean and these other tools develop, the tradeoffs may change.

% \mike{Maybe talk about model-based testing, a la Mathworks etc., here.}
% \awells{IMO we have better uses of space.}

\myparagraph{Differential and property-based random testing}
Perhaps the best-known example of \emph{differential testing} is the CSmith tool developed by~\citet{yang2011finding}, which tests C/C++ compilers against each other on randomly generated programs, looking for discrepancies in their results.
Other examples include~\citet{bornholt2021using}, who apply DRT to AWS S3's ShardStore, writing a simple model in Rust that serves as a test oracle and applying stateless model checking to prove properties of this simplified code.
\citet{groce2007randomized} use DRT as a precursor to formal methods, but they focus on correctness in the presence of hardware faults.
SybilFS~\cite{ridge2015sibylfs} proposes a reference model of a POSIX file system that other implementations can use as an oracle for differential testing.
%The authors note that SybilFS is amenable to formal proofs, but only prove two simple properties for an older version of their model.
%
% In general, differential testing has proved an effective method for finding bugs.

% \TODO{Change this paragraph to be about \emph{uses} of property testing for real systems, like the DropBox paper and also the Shardstore one (I think), rather than about quickcheck frameworks, as it is now.}
QuickCheck \cite{claessen2000quickcheck} introduced \emph{property-based random testing}, testing that a property holds on automatically-generated inputs, rather than on a few hand-defined ones.
Property testing is used in the S3 ShardStore paper mentioned above \cite{bornholt2021using}, in addition to model checking.
\citet{hughes2016mysteries} apply property testing to the distributed file systems Dropbox, Google Drive and ownCloud (an opensource equivalent) and found several bugs.
Property testing enjoys moderate popularity, e.g., the Hypothesis Python library~\cite{maciver2019hypothesis} has more than $200,000$ downloads per day as of January, 2024.
Defining a property and randomly testing it blurs the bounds between traditional testing and formal methods, and has been identified as a promising onramp to the use of formal methods~\cite{reid2020towards}.

\myparagraph{Dependability cases}
A \emph{dependability case}, as proposed by \citet{jackson2009direct}, is a careful collection of different sorts of evidence showing that a software system is correct.
Verification-guided development could be used to produce a dependability case: (1) evidence for good design is in the form of mechanized proofs of properties of the model, and (2) evidence of correct implementation is in the form of differential and property tests of the deployed code.
\citet{ernst2015toward} report that constructing a dependability case can lead to a clearer view of what assumptions underlie formal modeling and testing, which helps identify gaps to be shored up with further testing, proofs or other techniques.

\section{Conclusion}

This paper presented verification-guided development (VGD), a high-assurance engineering process that we use to develop the Cedar authorization language and tools.
The process has two parts.
\begin{enumerate}
    \item
We write a readable, executable model of Cedar in Lean and prove that the model satisfies key correctness and security properties.
Our proof effort leverages Lean's extensive theorem libraries, interactive IDE support, and fast verification.
\item We use differential random testing (DRT) to check that the Cedar production code, written in Rust, matches the model, and use property-based testing (PBT) to test properties against the production code for which there is no analogue in the model (or no proof, yet).
Our DRT/PBT input generators are carefully crafted to achieve good code coverage and balanced input distributions.
\end{enumerate}

Both proofs and DRT helped us to uncover and fix subtle bugs in various Cedar components prior to release.
Our experience shows that VGD is a practical approach for developing high-assurance code: it leverages the benefits of formal methods while producing code that is easy to use, develop, and maintain.\tighten

Cedar is open source: The Lean models, Rust code, and testing setup are all available at \url{https://github.com/cedar-policy}.

%\nocite{*}
\bibliography{bib}

% see these for the weird positioning:
% https://tex.stackexchange.com/questions/39017/how-to-influence-the-position-of-float-environments-like-figure-and-table-in-lat
% https://tex.stackexchange.com/questions/89462/page-wide-table-in-two-column-mode
\begin{table*}[t]
    \caption{Bugs found by DRT/PBT}
    \centering
    
    \begin{tabular}{|p{0.23\textwidth}|p{0.72\textwidth}|}\hline
        \textbf{Property} & \textbf{Description} \\\hline

Rust and Lean authorizer parity (ABAC) & The Rust implementation and definitional model computed string sizes differently \\\hline
Rust and Lean authorizer parity (ABAC type-directed) & The Rust implementation and definitional model named extension functions differently (e.g., \code{lt} vs. \code{lessThan}) \\\hline
Rust and Lean authorizer parity (ABAC type-directed) & The Rust implementation allowed octal numbers in IPv4 addresses while the definitional model did not \\\hline
Rust and Lean authorizer parity (ABAC type-directed) & The definitional model incorrectly rejected \code{"::"} when parsing IPv6 addresses \\\hline
Rust and Lean authorizer parity (ABAC type-directed) & The Rust implementation supported embedding IPv4 addresses in IPv6 addresses while the definitional model did not \\\hline
Rust and Lean authorizer parity (ABAC type-directed) & The Rust implementation required the operands of the \code{isInRange} extension function to be both IPv4 or IPv6 addresses while the definitional model did not \\\hline
Formatter roundtrip &  Comments on records were dropped by the formatter (\url{https://github.com/cedar-policy/cedar/pull/257}) \\\hline
Formatter roundtrip & Comments could be dropped in an \code{is} expression (\url{https://github.com/cedar-policy/cedar/pull/460}) \\\hline
Parser roundtrip & The parser did not unescape raw strings \\\hline
Parser roundtrip & The parser did not parse the pattern literals of the \code{like} operation correctly \\\hline
Parser roundtrip & The parser did not parse namespaced extension function names correctly \\\hline
Parser roundtrip & The parser performed constant folding incorrectly \\\hline
Parser roundtrip & Pretty-printing did not consistently add parentheses to negation operations \\\hline
Parser roundtrip & Pretty-printing certain function call ASTs resulted in a crash \\\hline
Rust and Lean validator parity & The Rust implementation and definitional model named extension functions differently (e.g., \code{ipaddr} vs. \code{IPAddr}) \\\hline
Rust and Lean validator parity & The Rust implementation incorrectly rejected certain policies with schemas containing unspecified entity types \\\hline
Rust and Lean validator parity & The Rust implementation incorrectly handled policies containing certain types of records (\url{https://github.com/cedar-policy/cedar/pull/165}) \\\hline
Rust and Lean validator parity & The Rust implementation and definitional model disagreed on the type of \code{resource in []} when \code{resource} has unspecified entity type (\url{https://github.com/cedar-policy/cedar/pull/615})  \\\hline
Validation soundness & The validator ignored certain entity type namespaces \\\hline
Validation soundness & The validator did not parse extension function call arguments correctly \\\hline
Validation soundness & The validator did not correctly typecheck certain \code{has} expressions \\\hline

    \end{tabular}
    
    \label{tab:trophy-cases}
\end{table*}

\newpage
\appendix
\section{Trophy and Anti-Trophy Cases}
\label{sec:appendix}

In this appendix, we list trophy (bugs found) and anti-trophy (bugs missed) cases of DRT and PBT, all of which have been promptly fixed by Cedar developers.
\Cref{tab:trophy-cases} lists the bugs found by Cedar's DRT/PBT. We annotate violations of the property \emph{Rust and Lean authorizer parity} with the generators used to find them.
\Cref{tab:anti-trophy-cases} lists missed bugs by Cedar's DRT/PBT.
We include links to the relevant pull requests when possible.
Items without links were fixed on earlier versions of Cedar, prior to open-sourcing.

\begin{table*}[t]
    \caption{Bugs missed by DRT/PBT}
    \centering
    \begin{tabular}{|p{0.47\textwidth}|p{0.11\textwidth}|p{0.35\textwidth}|}\hline
        \textbf{Description} & \textbf{Component} &\textbf{Root Cause} \\\hline

The Rust evaluator incorrectly implemented the \code{in} operation & Evaluator & DRT failed to generate inputs that trigger this bug\\\hline
The Rust evaluator accepted the string representation of an invalid decimal literal & Evaluator & Triggering input is too hard to generate\\\hline
The parser crashed on certain malformed policies & Policy parser & DRT does not methodically generate malformed policies\\\hline
The parser did not reject certain malformed policies (\url{https://github.com/cedar-policy/cedar/pull/594}) & Policy parser & DRT does not methodically generate malformed policies\\\hline
The API to link a policy to a template could crash on invalid inputs (\url{https://github.com/cedar-policy/cedar/pull/203}) & Public API & DRT did not test the relevant APIs\\\hline
Certain parsable policies could fail to be converted to their JSON representation (\url{https://github.com/cedar-policy/cedar/pull/601}) & Public API & DRT did not test the relevant APIs\\\hline
The JSON schema parser accepted inputs with unknown attributes & Schema parser & DRT does not test malformed schemas\\\hline
The validator did not terminate on certain inputs & Validator & Triggering input is too hard to generate\\\hline
The validator did not typecheck template-linked policies correctly (\url{https://github.com/cedar-policy/cedar/pull/371}) & Validator & DRT did not test the relevant APIs\\\hline

    \end{tabular}

    \label{tab:anti-trophy-cases}
\end{table*}
\end{document}